\pdfoutput=1
\documentclass{article}

\usepackage{graphicx,epsfig,color}
\usepackage{cite}
\usepackage[english]{babel}
\usepackage{amssymb,amsfonts,amsmath}
\usepackage{verbatim}
\usepackage{bbold}
\usepackage{hyperref}
\usepackage{subcaption}

\usepackage{stackrel}
\usepackage{tabularx}
\newcolumntype{Y}{>{\centering\arraybackslash}X}
\newcommand{\cM}{{\cal M}}
\newcommand{\cN}{{\cal N}}
\newcommand{\cO}{{\cal O}}
\newcommand{\cZ}{{\cal Z}}

\newcommand{\be}{\begin{equation}} \newcommand{\ee}{\end{equation}}
\newcommand{\bea}{\begin{eqnarray}} \newcommand{\eea}{\end{eqnarray}}
\newcommand{\beann}{\begin{eqnarray*}}  \newcommand{\eeann}{\end{eqnarray*}}
\newcommand{\bfig}{\begin{figure}} \newcommand{\efig}{\end{figure}}
\newcommand{\ba}{\begin{array}} \newcommand{\ea}{\end{array}}
\newcommand{\bcen}{\begin{center}} \newcommand{\ecen}{\end{center}}
\newcommand{\btab}{\begin{tabular}} \newcommand{\etab}{\end{tabular}}

\newcommand{\vev}[1]{\left\langle{#1}\right\rangle}

\usepackage[top=2.5cm, bottom=2.5cm, left=2.5cm, right=2.5cm]{geometry}	

\numberwithin{equation}{section}

\begin{document}

\begin{flushright}
HIP-2020-14/TH
\end{flushright}

\begin{center}

\centering{\Large {\bf Scattering length in holographic confining theories}}

\vspace{8mm}

\renewcommand\thefootnote{\mbox{$\fnsymbol{footnote}$}}
Carlos Hoyos,${}^{1,2}$\footnote{hoyoscarlos@uniovi.es}
Niko Jokela,${}^{3,4}$\footnote{niko.jokela@helsinki.fi} and
Daniel Logares${}^{1,2}$\footnote{dani.logares@gmail.com} 

\vspace{4mm}
${}^1${\small \sl Department of Physics} \\
{\small \sl Universidad de Oviedo}, {\small \sl c/ Federico Garc\'{\i}a Lorca 18, ES-33007 Oviedo, Spain} 

\vspace{2mm}
\vskip 0.2cm
${}^2${\small \sl Instituto Universitario de Ciencias y Tecnolog\'{\i}as Espaciales de Asturias (ICTEA)}\\
{\small \sl  Calle de la Independencia, 13, 33004 Oviedo, Spain}

\vspace{2mm}
\vskip 0.2cm
${}^3${\small \sl Department of Physics} and ${}^4${\small \sl Helsinki Institute of Physics} \\
{\small \sl P.O.Box 64} \\
{\small \sl FIN-00014 University of Helsinki, Finland} 

\end{center}

\vspace{0mm}

\renewcommand\thefootnote{\mbox{\arabic{footnote}}}

\begin{abstract}
\noindent  
The low-energy effective theory description of a confining theory, such as QCD, is constructed including local interactions between hadrons organized in a derivative expansion. This kind of approach also applies more generically to theories with a mass gap, once the relevant low energy degrees of freedom are identified. The strength of local interactions in the effective theory is determined by the low momentum expansion of scattering amplitudes, with the scattering length capturing the leading order. We compute the main contribution to the scattering length between two spin-zero particles in strongly coupled theories using the gauge/gravity duality. We study two different theories with a mass gap: a massive deformation of ${\cal N}=4$ super Yang-Mills theory (${\cal N}=1^*$) and a non-supersymmetric five-dimensional theory compactified on a circle. These cases have a different realization of the mass gap in the dual gravity description: the former is the well-known GPPZ singular solution and the latter a smooth $AdS_6$ soliton geometry. We find that the scattering lengths have similar functional dependences on the masses of the particles and on the conformal dimension of the operators that create them in both theories. Assuming these similarities hold more generally, they could be used to constrain the effective description of gapped strongly coupled theories beyond symmetry considerations.
\end{abstract}

\newpage

\tableofcontents


\newpage

\section{Introduction and summary}

There are many important physical systems whose complete description is out of reach of traditional approaches, typically because they are strongly coupled. Nevertheless, at long wavelengths and at low energies many details of microscopic physics become irrelevant and a simpler effective description is sufficient. A prominent example in the context of quantum chromodynamics (QCD) is the chiral effective theory. 

Bound with the effective theory does not mean that all the microscopic details are coarse-grained beyond reach. One can gain access to the microscopic physics as they are encoded in short-range interactions among the low energy degrees of freedom in the effective theory. It is even possible to do a systematic expansion of these interactions in terms of the wavelength over a microscopic length scale, the lowest order contribution being determined by the scattering length. This important text-book quantity can, in principle, be extracted from the zero momentum limit of the scattering amplitude involving the interacting degrees of freedom; for reviews on this approach, see \cite{Kaplan:2005es,Epelbaum:2008ga,Hammer:2016xye}.

However, a direct computation of the scattering amplitudes in the microscopic theory is typically not conceivable, especially at strong coupling. One either has to resort to experiments or take an indirect approach, for instance using numerical methods at finite volume \cite{Luscher:1986pf,Luscher:1990ux,Rummukainen:1995vs}. This also has some inherent limitations for heavier states in particular. A promising alternative is to use the holographic duality, which is well-suited for this task, as the gravity side becomes weakly coupled when the dual theory is strongly coupled.
 
In \cite{Hoyos:2019kzt} we proposed a method to compute scattering amplitudes and scattering lengths in theories with a mass gap and a discrete spectrum via holographic duality. We will dub these cases as ``confining'', even though we will not discuss the behavior of Wilson loops in this paper. We will work in the approximation of classical gravity, which in the dual field theory language translates into a type of large-$N$ expansion, with $N$ giving a measure of the number of degrees of freedom. The large-$N$ limit guarantees that the width of massive states is very small and thus allows us to consider asymptotic scattering processes. Scattering amplitudes will also be suppressed by $N$-dependent factors, thus comparison with numerical estimates or experiments requires an extrapolation from the large-$N$ limit.  

We illustrated the power of our method by finding the main contribution to the scattering length of a contact interaction in the two-to-two elastic scattering of spin-0 particles. We considered the hard wall model \cite{Erlich:2005qh}, consisting of a $AdS_5$ geometry with a sharp cutoff by imposing Dirichlet boundary conditions on the bulk fields. The position of the cutoff was identified as the scale of confinement, hence determining the masses of the particles in the field theory. As a model it is quite crude, and one might wonder whether the scattering amplitudes are very sensitive to the way the confinement scale is introduced. In other confining models the geometry ends smoothly, like in the Witten-Yang-Mills model \cite{Witten:1998zw}, or at a singularity, like in the dual to ${\cal N}=1^*$ super Yang-Mills (SYM) theory, the GPPZ solution \cite{Girardello:1999bd}. In principle, these differences could be reflected in the scattering amplitudes and lead to qualitatively different results. Our purpose in this work is to address this question by computing and comparing the scattering lengths in different classes of models. Note that the calculation of the scattering length is in a completely different regime from the hard scattering considered in some previous works \cite{Polchinski:2001tt}.

\begin{figure}[t!]
\begin{center}
\includegraphics[width=0.7\textwidth]{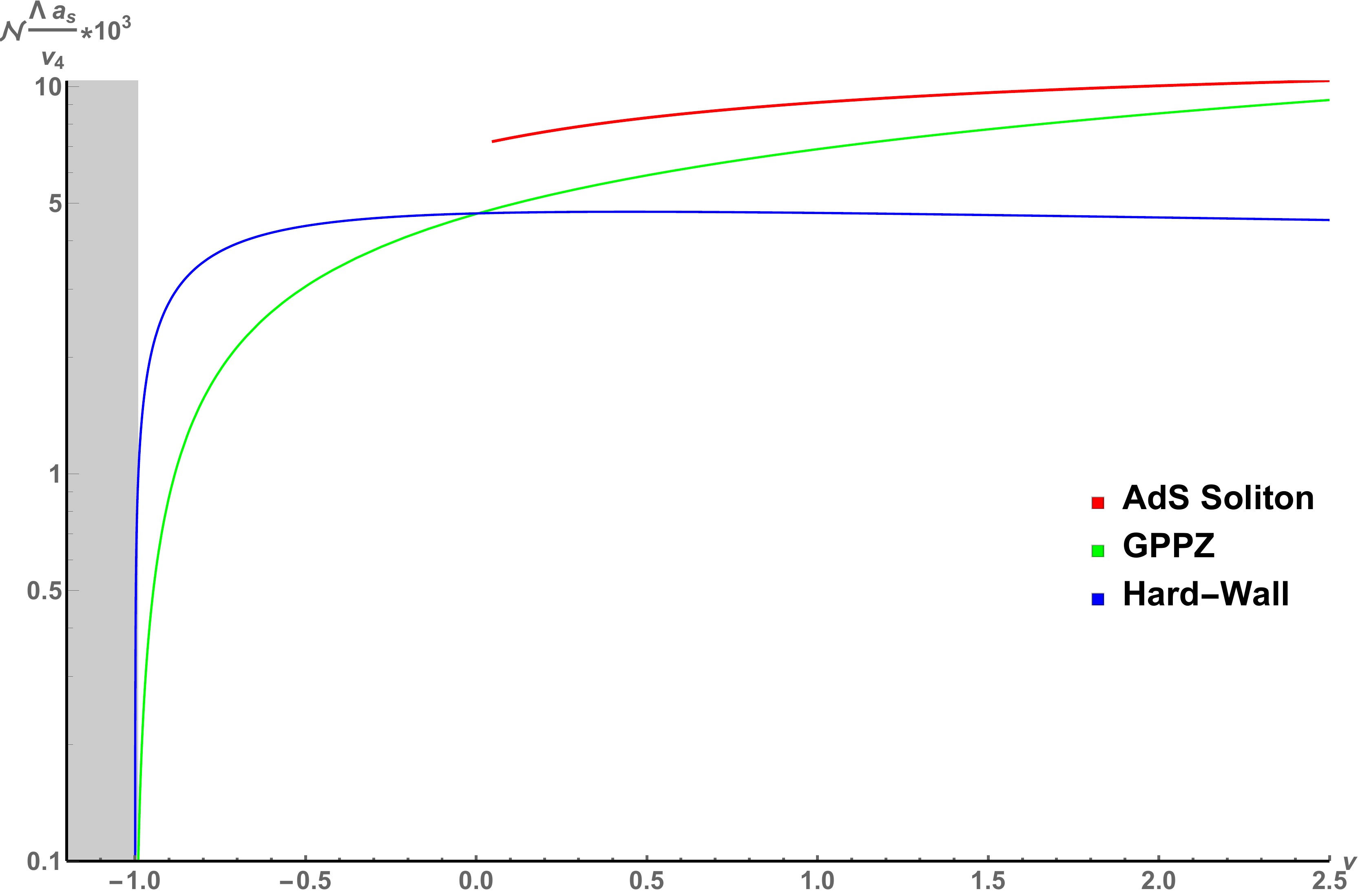}
\caption{Scattering length for the duals to the $AdS_6$ soliton, GPPZ geometry, and hard wall, in units of the confinement scale as a function the dimension of the dual operator $\nu=\Delta-\frac{d}{2}$. $\nu=-1$ corresponds to the unitarity bound, where the scattering length vanishes in all cases. The scattering length is roughly of the same order of magnitude and shows the same increasing trend with $\nu$ for the $AdS_6$ soliton and GPPZ model, while the hard wall model starts decreasing at larger values of $\nu$.}\label{fig:all}
\end{center}
\end{figure}

As two distinct representatives we consider the ${\cal N}=1^*$ SYM mentioned previously and a model similar to Witten-YM consisting of a non-supersymmetric $AdS_6$ soliton geometry, where one of the spatial directions collapses smoothly to zero size. Our results for the ${\cal N}=1^*$ SYM are represented in Fig.~\ref{fig:as0GPPZ}, corresponding to the scattering length for particles of the lowest masses as a function of the conformal dimension of the dual operator, and in Fig.~\ref{fig:asmGPPZ}, corresponding to the scattering length for fixed conformal dimension and different masses of the particles involved in the scattering process. The same quantities for the non-supersymmetric model are represented in Fig.~\ref{fig:as0sol} and in Fig.~\ref{fig:asmsol}.  They are qualitatively quite similar. There is a smooth increase of the scattering length with the conformal dimension along similar-looking curves. The scattering length is larger for scatterers of equal masses and decreases when any of the masses is increased. The conclusion is that the scattering length for this type of contribution is largely insensitive to the physics that produces the mass gap.\footnote{From a technical point of view the similarity probably originates in the solutions to the Sturm-Liouville problem in each geometry being similar.} For the hard wall model analogous figures can be found in \cite{Hoyos:2019kzt} (Figures 1, 2, and 3), there one can observe a similar behavior for the scattering length as a function of the masses. The dependence with the conformal dimension is also quite close, although in the hard wall model it is not monotonically increasing, suggesting that the hard wall model may miss some of the relevant physics for large-dimension operators. A direct comparison of the scattering lengths in each model can be seen in Fig.~\ref{fig:all}. It is, however, important to notice that while the functional dependences are very similar between the models of the present work, the $AdS_6$ soliton geometry always results in larger scattering lengths than in GPPZ and the quantitative match between the models is only within ${\cal{O}}(1)$. It is nevertheless tempting to argue that theories with a holographic dual always result in similar values for the scattering length together with a clear growing trend with low operator dimension. We leave this as an interesting open problem to be investigated in future works, however.

The dependence on the mass can be partially understood. This is because the scattering amplitude is proportional to the overlap between the modes in the gravity dual, which is bound to decrease as the masses become more apart, and the modes will have support on different regions. In field theory this would imply that particles of very different masses have weaker contact interactions. Although we do not have a clear cut explanation, a possible interpretation is that the particles with very different Compton wavelengths are less likely to scatter: imagine the one with smaller wavelength as a particle-like object of the size of its Compton wavelength and the other as a wave.

As a function of the conformal dimension, we observe in the ${\cal N}=1^*$ SYM and hard wall models that the scattering length vanishes when the unitarity bound is saturated, conforming to expectations, as this point corresponds to free particles. It is then natural that the scattering becomes stronger as the dimension increases above the unitarity bound.  For the $AdS_6$ soliton we did not reach the unitarity bound but indeed we observe that the scattering length increases with the conformal dimension. The overall sign of the scattering length depends on whether the interaction is repulsive (positive) or attractive (negative). In the plots, this sign is determined by the factor $v_4$ that defines the contact interaction in the gravity dual.  

The rest of this paper is organized as follows. We will start by reviewing in Sec.~\ref{sec:setup} the main ingredients leading to the scattering length. We then continue in Secs.~\ref{sec:GPPZ} and \ref{sec:AdS6} with the analysis of $\cN=1^*$ SYM and non-supersymmetric model, respectively. We finish commenting on some possible future directions in Sec.~\ref{sec:outlook}. Several technical details concerning analytic and numerical calculations are collected in the Appendices following the main text.

\section{Scattering length in holographic theories}\label{sec:setup}

In this section we present the salient features of \cite{Hoyos:2019kzt} to compute scattering amplitudes in $d$-dimensional strongly coupled theories with a $(d+1)$-dimensional holographic dual. In holographic models that are dual to a confining theory there is a discrete spectrum of normalizable modes that map to particles, glueballs, or mesons in the field theory dual. Scattering between these particles can be obtained by applying LSZ reduction formulas to correlators of gauge-invariant operators with the quantum numbers of the particles involved. In the holographic calculation this implies that it is enough to identify the leading pole contributions in bulk correlators when the momenta are taken on-shell. The corresponding residues then determine the scattering amplitudes.

A (real-valued) gauge-invariant single-trace local operator will map to a scalar field $\Phi$ in the holographic gravity dual. In this paper we will restrict to cases where the scalar field is treated as a probe, so backreaction on the metric and on other fields will be neglected. The generic bulk action will have the form
\be\label{eq:action}
 S=-\frac{1}{8\pi G_N}\int d^{d+1}x\,\sqrt{-g}\left(\frac{1}{2} g^{MN}\partial_M\Phi\partial_N\Phi+V(\Phi)\right) \ .
\ee
Here $g_{MN}$, $M,N=0,1,\ldots,d$, is the $d+1$ dimensional metric, that we will assume is asymptotically $AdS_{d+1}$ with radius $L$
\be\label{eq:adsmetric}
 ds_{d+1}^2=g_{MN}dx^M dx^N \underset{z\to 0}{\longrightarrow} \frac{L^2}{z^2}\left( dz^2+\eta_{\mu\nu}dx^\mu dx^\nu\right) \ .
\ee
In these coordinates $x^\mu$, $\mu=0,1,\ldots,d-1$ span the directions of the field theory dual.

We will consider scalar potentials admitting an analytic expansion for small values of the field $\Phi$:
\be\label{eq:potential}
 V(\Phi)=\frac{1}{2}m^2 \Phi^2+\frac{v_4}{2L^2}\Phi^4+\ldots \ ,
\ee
where the mass $m$ is determined by the conformal dimension of the dual operator through the usual relation
\be
 m^2L^2=\Delta(\Delta-d) \ .
\ee
In the following, it is useful to parametrize the conformal dimension as
\be
 \Delta=\frac{d}{2}+\nu\ ,
\ee
where $-1<\nu<d/2$ is the range of allowed values for relevant operators satisfying the unitarity bound.

\subsection{Solutions and propagators of the scalar field}

The linearized equation of motion for the scalar field is
\be
 (\square -m^2)\Phi=\frac{1}{\sqrt{-g}}\partial_M\left( \sqrt{-g} g^{MN}\partial_N\Phi\right)-m^2 \Phi=0\ .
\ee
Solutions to the equation of motion stemming from (\ref{eq:action}) can be constructed perturbatively using the bulk-to-boundary $K(x,x';z)$ and bulk-to-bulk $G(x,x';z,z')$ propagators by including higher order terms of the potential (\ref{eq:potential}). The bulk-to-boundary propagator determines the linearized solution for any boundary condition
\be
 (\square_{z,x}-m^2)K=0\ , \ K(x,x';z)\underset{z\to 0}{\longrightarrow} z^{d-\Delta} \delta^{(d)}(x-x')\ .
\ee
The bulk-to-bulk propagator on the other hand is the Green's function defined by the differential equation 
\be
(\square_{z,x}-m^2)G=\frac{1}{\sqrt{-g}}\delta^{(d)}(x-x')\delta(z-z')\ , \ G(x,x';z,z')\underset{z\to 0}{\longrightarrow} z^\Delta \ .
\ee

In order to accommodate the $AdS$ soliton model that we study later, we will assume Poincar\'e invariance of the full geometry along a subset of the field theory directions $d_{eff}\leq d$ and remove all explicit dependence from the remaining field theory directions.  We will thus split the metric as follows
\be
 ds_{d+1}^2=g_{zz}dz^2+g_{xx}\eta_{\mu\nu}dx^\mu dx^\nu+g_{ij}dy^i d y^j \ ,
\ee
where now $x^\mu$, $\mu=0,1,\ldots, d_{eff}-1$ and $y^i$, $i=d_{eff},\ldots,d-1$. Written in this way, the plane spanned by the vectors $x^\mu$ retain the Poincar\'e symmetry, while directions $y^i$ are orthogonal to this plane. We proceed by expanding the scalar field and the propagators in plane waves
\be
 \Phi(x,z)=\int \frac{d^{d_{eff}}p}{(2\pi)^{d_{eff}}}\phi(p,z)e^{ip\cdot x} \ .
\ee
For a confining theory, there will be a discrete set of normal modes satisfying conditions of regularity at the origin and being normalizable at the boundary
\be
p^2=-M_n^2 \ , \ \phi=\varphi_n(z) \ , \ n=0,1,2,3,\ldots \ . 
\ee
The spectrum of masses $M_n^2$ corresponds to the spectrum of massive states in the dual field theory associated to the dual scalar operator. The equation of motion for the scalar can be put in Sturm-Liouville form
\be
 \partial_z\left(\sqrt{-g} g^{zz}\partial_z\phi\right)-m^2\sqrt{-g}\phi+\sqrt{-g}g^{xx}M^2\phi=0 \ .
\ee
We can identify $\rho(z)=\sqrt{-g}g^{xx}$ with the weight, so that normal modes can be chosen to form an orthonormal basis
\be
 \int dz\, \rho(z) \varphi_n(z)\varphi_m(z)=\delta_{nm} \ .
\ee
The integration is over the whole allowed range of the radial coordinate. The basis of normal modes can be used to write an expression for the bulk-to-bulk propagator
\be\label{eq:Gprop}
 G(p;z,z')=-\sum_n \frac{\varphi_n(z)\varphi_n(z')}{p^2+M_n^2} \ .
\ee

\subsection{Residues and quartic vertex contribution}

The quartic term in the scalar potential gives a contribution to the scattering amplitude that can be interpreted as being produced by a Witten diagram with four scalar legs joining at a single point in the bulk. The relevant quantity is, omitting a trivial factor imposing energy-momentum conservation,
\be
 G^{(3)}(z,p_1;-p_2,-p_3,-p_4)\propto \, v_4 \int dz'\, \sqrt{-g} G(p_1;z,z')K(p_2;z')K(p_3;z')K(p_4;z') \ .
\ee
When the momenta are taken on-shell, both bulk-to-boundary and bulk-to-bulk propagators have poles
\be\label{eq:poles}
 K(p;z)\xrightarrow[p^2\to -M_n^2]{} \frac{c_n \varphi_n(z)}{p^2+M_n^2}\ , \ G(p;z,z')\xrightarrow[p^2\to -M_n^2]{} -\frac{\varphi_n(z)\varphi_n(z')}{p^2+M_n^2}\ .
\ee
Close to the boundary, the residue of the boundary-to-bulk propagator has the form
\be
 c_n \varphi_n(z) \underset{z\to 0}{\longrightarrow} \frac{Z_n}{2\nu\cN}\; z^\Delta\ ,
\ee
where $\cN=L^{d-1}/(16\pi G_N)$ is a dimensionless normalization factor proportional to the number of degrees of freedom in the dual field theory. The factor $Z_n$ can be identified with the residue of the massive pole in the two-point function of the dual scalar operator, explicitly
\be\label{eq:res}
 Z_n=2\nu\cN \, c_n\times \lim_{z\to 0}\frac{\varphi_n(z)}{z^\Delta} \ .
\ee

When all the momenta are taken on-shell, the leading pole contribution to the amplitude is
\be
 G^{(3)}(z,p_1;-p_2,-p_3,-p_4) \xrightarrow[p_i^2\to -M_{n_i}^2]{} \frac{\Gamma^{(3)}_{n_1; n_2,n_3,n_4}(z)}{\prod_{i=1}^4 (p_i^2+M_{n_i}^2)}\ .
\ee
The boundary expansion of the residue takes the form
\be
\Gamma^{(3)}_{n_1; n_2,n_3,n_4}(z)\underset{z\to 0}{\longrightarrow} \frac{\cZ_{n_1; n_2,n_3,n_4}}{2\nu \cN}\; z^\Delta\ . 
\ee
The factor $\cZ_{n_1; n_2,n_3,n_4}$ is the leading pole contribution to the four-point function of the dual scalar operator. Through the LSZ reduction formula it will determine the scattering amplitude. Its explicit form is
\be\label{eq:res4}
\cZ_{n_1; n_2,n_3,n_4}=-2\nu v_4\, \cN \,c_{n_2}c_{n_3}c_{n_4} \kappa_{_{n_1},_{n_2},_{n_3},_{n_4}}\times \lim_{z\to 0}\frac{\varphi_{n_1}(z)}{z^\Delta} \ ,
\ee
where the overlap $\kappa$ is defined as
\be
 \kappa_{_{n_1},_{n_2},_{n_3},_{n_4}} =\int dz' \sqrt{-g} \prod_{i=1}^4 \varphi_{n_i}(z')\ .
\ee
With the residues \eqref{eq:res} and \eqref{eq:res4}, the contribution of the quartic term to the scattering amplitude is
\be\label{eq:amplit}
\cM_{n_1,n_2,n_3,n_4}=\frac{1}{4}\sum_{i=1}^4 \frac{\cZ_{n_i;  \{ n_k\neq n_i\}}}{(Z_{n_1}Z_{n_2}Z_{n_3}Z_{n_4})^{1/2}}\ .
\ee
If $d_{eff}=4$, the scattering length for two-to-two elastic scattering can be determined directly from the previous amplitude using the formula derived in \cite{Hoyos:2019kzt}
\be\label{eq:as}
a_s=-\frac{\cM_{n_1=n_3,n_2=n_4}}{8\pi(M_{n_1}+M_{n_2})} \ .
\ee

\section{Scattering in the dual to $\cN=1^*$ super Yang-Mills}\label{sec:GPPZ}

The $\cN=1^*$ SYM theory we will study is a massive deformation of $\cN=4$ SYM whose gravity dual was found by Girardello, Petrini, Porrati, and Zaffaroni \cite{Girardello:1999bd} and is commonly known as the GPPZ solution. In the field theory side, the matter content of $\cN=4$ SYM is split in a $\cN=1$ vector multiplet and three chiral multiplets in the adjoint representation. An equal mass is given for the three chiral superfields, in such a way that the global symmetry group is broken to $SU(3)$ and supersymmetry is broken to $\cN=1$. At weak coupling the theory flows to pure $\cN=1$ SYM at energy scales much below the mass of the chiral multiplets. The weakly coupled theory is confining and this property holds in the strongly coupled theory, as two-point functions of gauge-invariant operators show poles for a discrete spectrum of massive states \cite{Anselmi:2000fu,Papadimitriou:2004rz}. 

We will not study the most general case, but restrict to the simpler subset of vanishing gaugino condensate in this paper. In this case the background geometry is a a solution of five-dimensional supergravity truncated to a single scalar coupled to the metric. In a convenient set of coordinates the metric is ($0< u <1$)
\be
 ds^2_{4+1}=L^2\left(\frac{du^2}{4(1-u)^2}+\frac{u}{1-u}\eta_{\mu\nu}dx^\mu dx^\nu\right) \ .
\ee
The metric can be put in the form given in \eqref{eq:adsmetric} by a change of coordinates 
\be
 u=1-\frac{z^2}{z_\Lambda^2}, \ \ x^\mu \to  \frac{1}{z_\Lambda} x^\mu \ .
\ee
Therefore, $u\to 1$ corresponds to the boundary of the bulk, which is asymptotically $AdS_5$. The other limit $u\to 0$ is the origin of bulk spacetime $z\to z_\Lambda$. The scale of confinement is $\Lambda=1/z_\Lambda$; in the following we will set $z_\Lambda=1$, so all dimensionful quantities are given in units of $\Lambda$.

\subsection{Scalar solutions}

Although there are several scalar fields in the supergravity action, their action involves coupling the background metric to the scalar field that will subsequently make the analysis more involved. This will affect the spectrum of normal modes and moreover introduce cubic couplings \cite{Bianchi:2003bd,Bianchi:2003ug}. These technical complications may obscure the physics we are primarily interested in, which is how the geometric realization of confinement affect to the scattering amplitudes. Thus, in order to facilitate the analysis and the comparison with other models, we will study a family of probe scalar fields, decoupled from the background scalar and with a quartic potential as presented in the general analysis.

A scalar operator of dimension $\Delta=2+\nu$ is dual to a scalar field of mass 
\be
 m^2L^2=\Delta(\Delta-4)=\nu^2-4 \ .
\ee
The linearized equation of motion for the scalar field is
\be\label{eq:eomGPPZ}
 \phi''+\frac{2-u}{1-u}\phi' +\left(\frac{M^2}{4(1-u)}-(\nu^2-4)\frac{u}{4(1-u)^2}\right)\phi=0\ .
\ee
Regular solutions are given in terms of hypergeometric functions
\be\label{eq:phiGPPZ}
 \phi_M(u)=(1-u)^{\frac{2+\nu}{2}}\,_{2}F_{1}\left(1+\frac{\nu}{2}-\frac{1}{2}\sqrt{\nu^{2}+M^{2}-4},1+\frac{\nu}{2}+\frac{1}{2}\sqrt{\nu^{2}+M^{2}-4};2;u\right) \ .
\ee
The bulk-to-boundary propagator $K_\nu(M;u)$ is proportional to (\ref{eq:phiGPPZ}), normalized to have the right asymptotic behavior
\be
 K_M(u) \underset{u\to 1}{\longrightarrow}  1\cdot  (1-u)^{\frac{2-\nu}{2}}= z^{2-\nu} \ .
\ee
Then,
\be\label{eq:KGPPZ}
 K_M(u)=\frac{\Gamma\left(1-\nu\right)}{\pi\csc\left(\pi\nu\right)}\Gamma\left(1+\frac{\nu}{2}-\frac{1}{2}\sqrt{\nu^{2}+M^{2}-4}\right)\Gamma\left(1+\frac{\nu}{2}+\frac{1}{2}\sqrt{\nu^{2}+M^{2}-4}\right)\phi_M(u)\ .
\ee
Poles in the bulk-to-boundary propagator correspond to the spectrum of normalizable modes. In this case poles appear when the argument of the middle $\Gamma$ function in (\ref{eq:KGPPZ}) is a non-positive integer
\be
 1+\frac{\nu}{2}-\frac{1}{2}\sqrt{\nu^{2}+M^{2}-4}=-n,\ \ n=0,1,2,\ldots \ .
\ee
This gives the following mass spectrum
\be\label{eq:MnGPPZ}
 M_n^2= 4(n^2 +(2+\nu)n+2+\nu) \ .
\ee
At these values the scalar solutions \eqref{eq:phiGPPZ} become
\be\label{eq:normalphi}
 \phi_{M_n}(u)=(1-u)^{\frac{2+\nu}{2}}\,_{2}F_{1}(-n,n+\nu+2;2;u)\ .
\ee
The normal modes are $\varphi_n=\alpha_n \phi_{M_n}(u)$, where
\be
 \alpha_n=\sqrt{2\left(n+1\right)\left(\nu+n+1\right)\left(\nu+2n+2\right)}\ .
\ee
They form an orthonormal basis respect to the weight $\rho(u)=u/(2(1-u)^2)$
\be
 \int_0^1 du\, \rho(u) \varphi_n(u)\varphi_m(u)=\delta_{nm} \ .
\ee
The computation is straightforward and we have relegated the details in Appendix~\ref{app:gppz}. The bulk-to-bulk propagator is determined via \eqref{eq:Gprop}.

\subsection{Scattering for low mass states}

\begin{figure}[t!]
\caption{Scattering length in $\cN=1^*$ SYM}
\begin{subfigure}{0.45\textwidth}
\includegraphics[width=\textwidth]{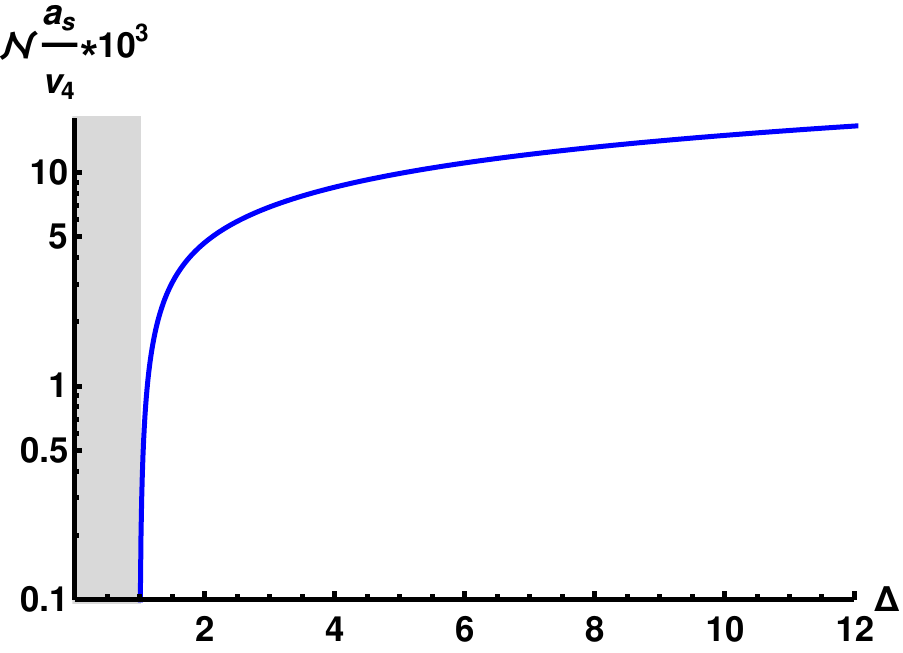}
\caption{Rescaled scattering length in units of the confinement scale. The horizontal axis is the conformal dimension of the scalar operator that creates the particles involved in the scattering, in this case the scattering is for the particles of lowest mass at each value of the conformal dimension. Notice the log-linear scale.}\label{fig:as0GPPZ}
\end{subfigure}
\hspace{8pt}
\begin{subfigure}{0.45\textwidth}
\includegraphics[width=\textwidth]{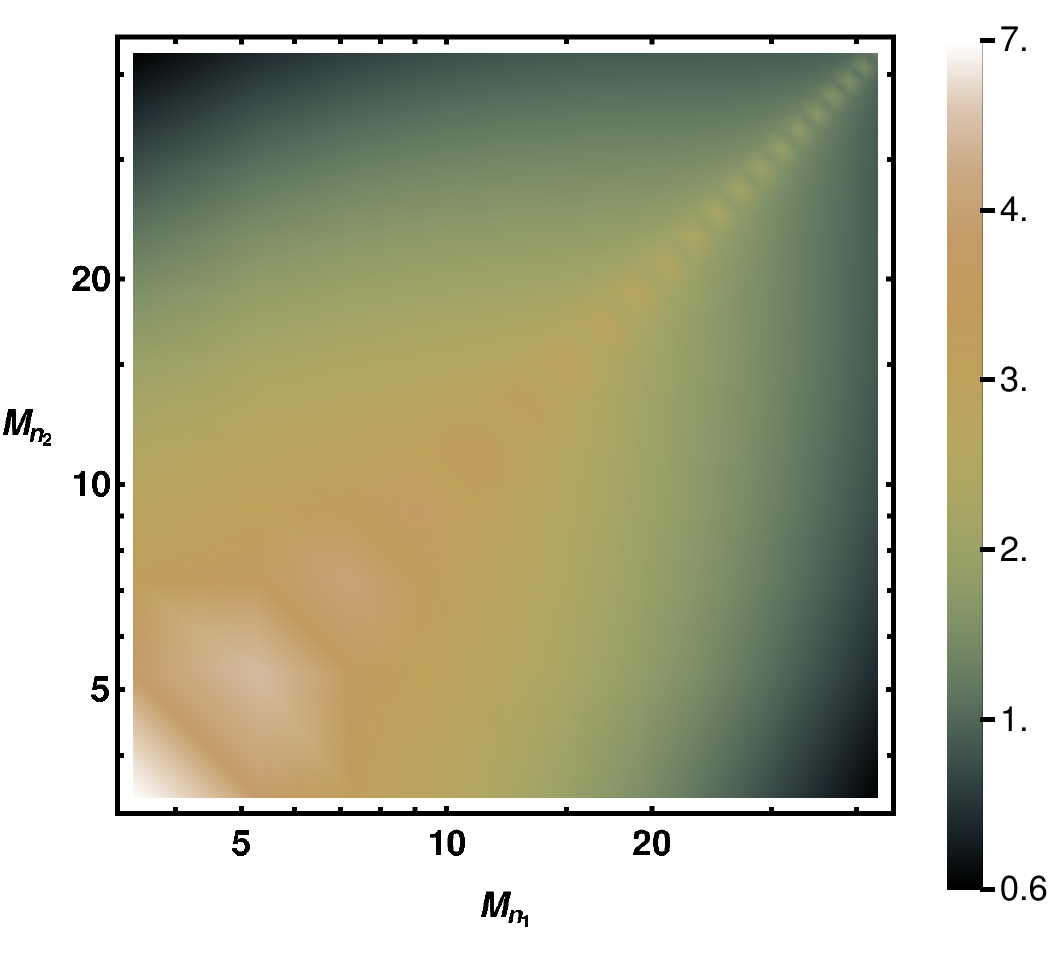}
\caption{Density plot of the scattering length $10^3\times {\cal{N}} a_s/v_4$ in units of the confinement scale. The vertical and horizontal axes scale logarithmically and indicate the masses of the particles involved in the scattering, in units of the confinement scale, for states created by a scalar operator of fixed conformal dimension $\Delta=3$ ($\nu=1$).}\label{fig:asmGPPZ}
\end{subfigure}
\end{figure}

Using the values \eqref{eq:MnGPPZ}, the leading pole in the bulk-to-boundary propagator \eqref{eq:KGPPZ} is of the form \eqref{eq:poles}, with 
\be
c_n=\frac{4(2+\nu+2n)(-1)^n}{n!\alpha_n}\frac{\Gamma\left(1-\nu\right)}{\pi\csc\left(\pi\nu\right)}\Gamma\left(2+\nu+n\right)\ .
\ee
Therefore, the residue of the two-point function \eqref{eq:res} is
\be
Z_n=2\nu \cN \,c_n \alpha_n \,_{2}F_{1}(-n,n+\nu+2;2;1)\ .
\ee
For the lowest mass $n=0$, we have that
\be
\alpha_0=\sqrt{2(\nu+1)(\nu+2)}\ ,  \ c_0=4\nu(\nu+1)(\nu+2)/\alpha_0\ ,
\ee
and the residue is
\be\label{eq:Z2GPPZ}
 Z_0=8\cN \nu^2(\nu+1)(\nu+2)\ .
\ee

The residue of the leading pole in the four-point function is given by \eqref{eq:res4}, 
\be
 \cZ_{n_1;n_2,n_3,n_4}=-2\nu v_4\, \cN \, c_{n_2} c_{n_3} c_{n_4}  \kappa_{_{n_1},_{n_2},_{n_3},_{n_4}}\, \alpha_{n_1} \,_{2}F_{1}(-n_1,n_1+\nu+2;2;1)\ ,
\ee
where the overlap is
\be
 \kappa_{_{n_1},_{n_2},_{n_3},_{n_4}} =\int_0^1 du\,\frac{u^2}{2(1-u)^3} \prod_{i=1}^4 \varphi_{n_i}(u)\ .
\ee
If all the scatterers have equal masses $M_0$, then this simplifies to
\be
 \kappa_{0,0,0,0} =\int_0^1 du\,\frac{u^2}{2(1-u)^3} \varphi_0^4(u)=\frac{\alpha_0^4}{2} \int_0^1 du\,u^2 (1-u)^{1+2\nu}=\frac{\alpha_0^4}{4(\nu+1)(\nu+2)(2\nu+3)}\ .
\ee
Then,
\be\label{eq:Z4GPPZ}
 \cZ_{0;0,0,0}=-64 \cN\, v_4\,  \frac{\nu ^4 (\nu +1)^3 (\nu +2)^3}{2\nu+3}\ .
\ee

Let us proceed with our calculation to extract the scattering amplitudes for equal mass lowest modes. For this it is enough to use the formulas \eqref{eq:as} and \eqref{eq:amplit} by plugging in the values for the residues \eqref{eq:Z2GPPZ} and \eqref{eq:Z4GPPZ}. We therefore find the amplitude
\be
\cM_{0,0,0,0}=-\frac{v_4}{\cN}\frac{(\nu +1) (\nu +2)}{2\nu+3}\ ,
\ee 
and the scattering length
\be\label{eq:GPPZas}
 a_s=\frac{v_4}{\cN}\frac{(\nu+1)(\nu+2)^{1/2}}{32\pi  (2\nu+3) }\ .
\ee
As shown in Appendix~\ref{app:gppz}, this expression is also valid for $\Delta<\frac{d}{2}$, extrapolating it to values $-1<\nu<0$.  We have represented the result for the scattering length in Fig.~\ref{fig:as0GPPZ}. 
We have also computed the scattering length for $\nu=1$ and scatterers of different mass, the results are represented in Fig.~\ref{fig:asmGPPZ}. 

\section{Scattering in non-supersymmetric theories}\label{sec:AdS6}

Having discussed scattering in a concrete confining background geometry, it is interesting to raise the question of how universal the results are and, in particular, how sensitive the scattering length is quantitatively to different conformal symmetry breaking mechanisms. To this end, we will study a different type of confining theory, obtained through supersymmetry-breaking compactifications. At strong coupling there is no separation between the confinement and the compactification scales, so glueballs and Kaluza-Klein modes have similar masses. Nevertheless, the theory is effectively four-dimensional with a mass gap, so it is still meaningful to compute the scattering amplitudes among different massive states.

To be more precise, we will take as background geometry $AdS_6$ compactified along one direction. This should be the dual geometry to a five-dimensional conformal field theory compactified along one direction with supersymmetry-breaking boundary conditions. The geometry is the $AdS$ soliton, that can be obtained by a double Wick rotation of a black brane solution \cite{Horowitz:1998ha}
\be
 ds^2=\frac{L^2}{z^2}\left( \frac{dz^2}{f(z)}+f(z)d\tau^2+\eta_{\mu\nu}dx^\mu dx^\nu\right),\ \ f(z)=1-\frac{z^5}{z_\Lambda^5}\ .
\ee
Here $\tau$ is the compact coordinate $\tau\sim \tau + 2\pi/M_{KK}$, with $M_{KK}=5/(2 z_\Lambda)$ the compactification scale. The space ends at a finite value of the radial coordinate $z=z_\Lambda$. As in previous examples, we identify $\Lambda=1/z_\Lambda$ with the scale of confinement. While we will not discuss an explicit holographic dual field theory, this geometry can in principle be embedded in string theory, as deformations of $AdS_6$ solutions of Type II supergravity \cite{Apruzzi:2014qva,Kim:2015hya}.

The $AdS_6$ soliton geometry is a close cousin of  the Witten-YM model \cite{Witten:1998zw}, extensively used to mimic QCD in applications of holography. The model consists of a non-supersymmetric compactification of D4-branes along a circle, and becomes pure Yang-Mills at low energies and weak coupling. The holographic dual geometry is also the $AdS_6$ soliton in an appropriately chosen frame, but in addition there is background dilaton, to which other fields may be coupled \cite{Kanitscheider:2008kd}. Alternatively, the Witten-YM model can be obtained from the compactification of $M5$-branes along a two-torus. In the holographic dual description the geometry is $AdS_7$ compactified along two spatial directions.

\subsection{Normalizable solutions for probe scalar fields}

A scalar operator of dimension $\Delta=\frac{5}{2}+\nu$ is dual to a scalar field of mass 
\be\label{eq:scala}
 m^2L^2=\Delta(\Delta-5)=\nu^2-\frac{25}{4}\ .
\ee
Note that $\Delta$ is the conformal dimension in the five-dimensional CFT. In the effective four-dimensional field theory the effective dimension of the dual operator will in general be different. Let us proceed with computing the spectrum in the following. The spectra of fluctuations in $AdS_6$ soliton geometries has been worked out previously in \cite{Wen:2004qh,Kuperstein:2004yf,Elander:2018aub} as well as in \cite{Elander:2013jqa,Elander:2020csd}, where interesting extensions are investigated.
However, as in previous section, we will focus on probe scalars and their spectra for arbitrary conformal dimension as represented in (\ref{eq:scala}). 

The linearized equation of motion for the scalar field is
\be\label{eq:EOMsol}
 \phi''(z)+\left(\frac{f'}{f}-\frac{4}{z} \right)\phi'(z)+\left(\frac{M^2}{f}-\frac{\nu^2-\frac{25}{4}}{z^2f}\right)\phi(z)=0\ .
\ee
The Sturm-Liouville form of this equation gives a weight $\rho(z)=1/z^4$. We do not know an analytic solution to this equation, so we will resort to a numerical calculation to compute the scattering amplitude and the scattering length. Close to the $AdS_6$ boundary, the scalar field has a leading order behavior $\phi \sim z^{\frac{5}{2}-\nu}$. We will factor out this dependence so that the scalar solution goes to a constant at the boundary. Defining $\phi=z^{\frac{5}{2}-\nu} \chi$ the equation of motion becomes
\be
 \chi''+\left(\frac{f'}{f}+\frac{1-2\nu}{z} \right)\chi'(z)+\left(\frac{M^2}{f}+\left(\frac{5}{2}-\nu\right)\frac{f'}{zf}+\frac{\nu^2-\frac{25}{4}}{z^2}\left(1-\frac{1}{f}\right) \right)\chi(z)=0\ .
\ee
As before, we will fix $z_\Lambda=1$ and express all quantities in units of $\Lambda$. We then impose regularity at the origin $z=1$ and do numerical shooting to the boundary. The details about the numerical calculation can be found in Appendix~\ref{app:sol}.

\begin{figure}[t!]
\begin{center}
\includegraphics[width=0.7\textwidth]{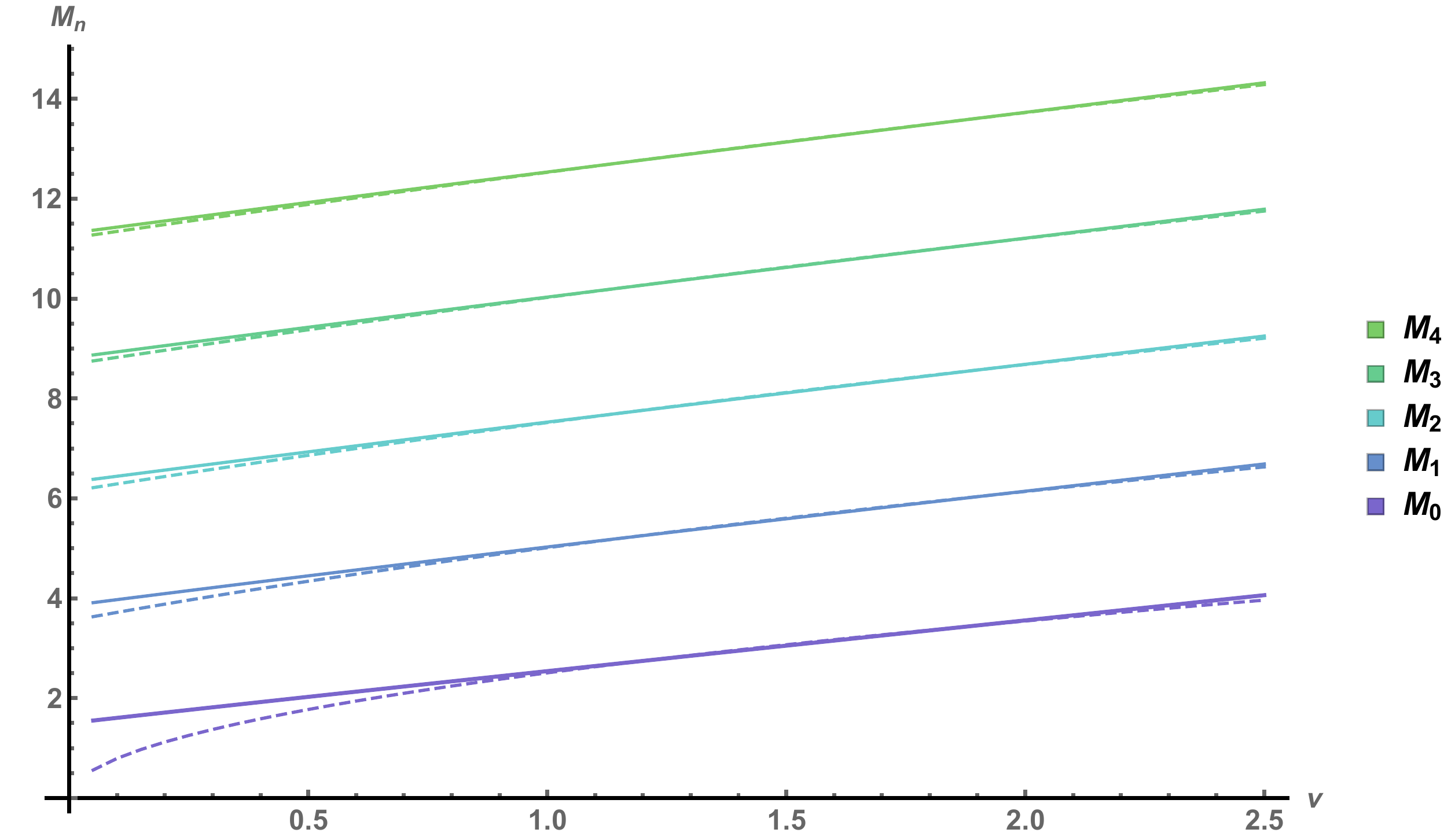}
\caption{Masses of the lower modes in units of the confinement scale. The solid curves correspond to the numerical results while the dashed curve stem from the WKB approximation obtained in Appendix~\ref{app:wkb}. }\label{fig:masses}
\end{center}
\end{figure}

For a given mass $M$, if the value of the solution $\chi_{_M}$ at the boundary is non-zero $\chi_{_M}(0)\neq 0$, the bulk-to-boundary propagator can be defined as
\be\label{eq:Ksol}
 K_M(z)=z^{\frac{5}{2}-\nu}\frac{\chi_{_M}(z)}{\chi_{_M}(0)} \ .
\ee
The spectrum of normal modes is determined by the values $M_n$ for which $\chi_{_{M_n}}(0)=0$, which can be found numerically; see Appendix~\ref{app:sol}. We have also derived an analytic estimate for the masses using a WKB approximation; see Appendix~\ref{app:wkb}
\be
 M_n^2=\frac{\pi^2}{\xi^2}(n+1)(n+\nu)+O(n^0) \ , \ n\geq 0 \ ,
\ee
where $\xi\approx 1.25$. For GPPZ we obtain that the difference between the exact and WKB result is $\nu$- and $n$- independent $M^2_{\rm exact}-M^2_{WKB}=4$. Comparison between the WKB values and numerical values in the $AdS_6$ soliton show also very good agreement, especially as the value of $\nu$ is increased. The spectrum for the first few modes and different values of $\nu$ is represented in Fig.~\ref{fig:masses}.

The fully normalizable solutions are given by
\be
 \varphi_n(z) \equiv \alpha_n \phi_{n}(z)=\alpha_n\, z^{\frac{5}{2}-\nu}\chi_{_{M_n}}(z)\ , 
\ee
where the coefficients $\alpha_n$ are chosen in such a way that the normal modes have unit norm, forming a complete orthonormal set
\be
 \int_0^1 \frac{dz}{z^4} \varphi_n(z)\varphi_m(z)=\delta_{nm} \ , \ \alpha_n = \left(\int_0^1 \frac{dz}{z^4}\phi_n(z)^2\right)^{-1/2} \ .
\ee
Some numerical values of the normalization coefficients are given in Appendix~\ref{app:sol}. The bulk-to-bulk propagator is determined by this basis as in \eqref{eq:Gprop}.

\subsection{Scattering for low mass states}

\begin{figure}[t!]
\begin{center}
\caption{Scattering length in the non-supersymmetric holographic model}
\begin{subfigure}{0.45\textwidth}
\includegraphics[width=\textwidth]{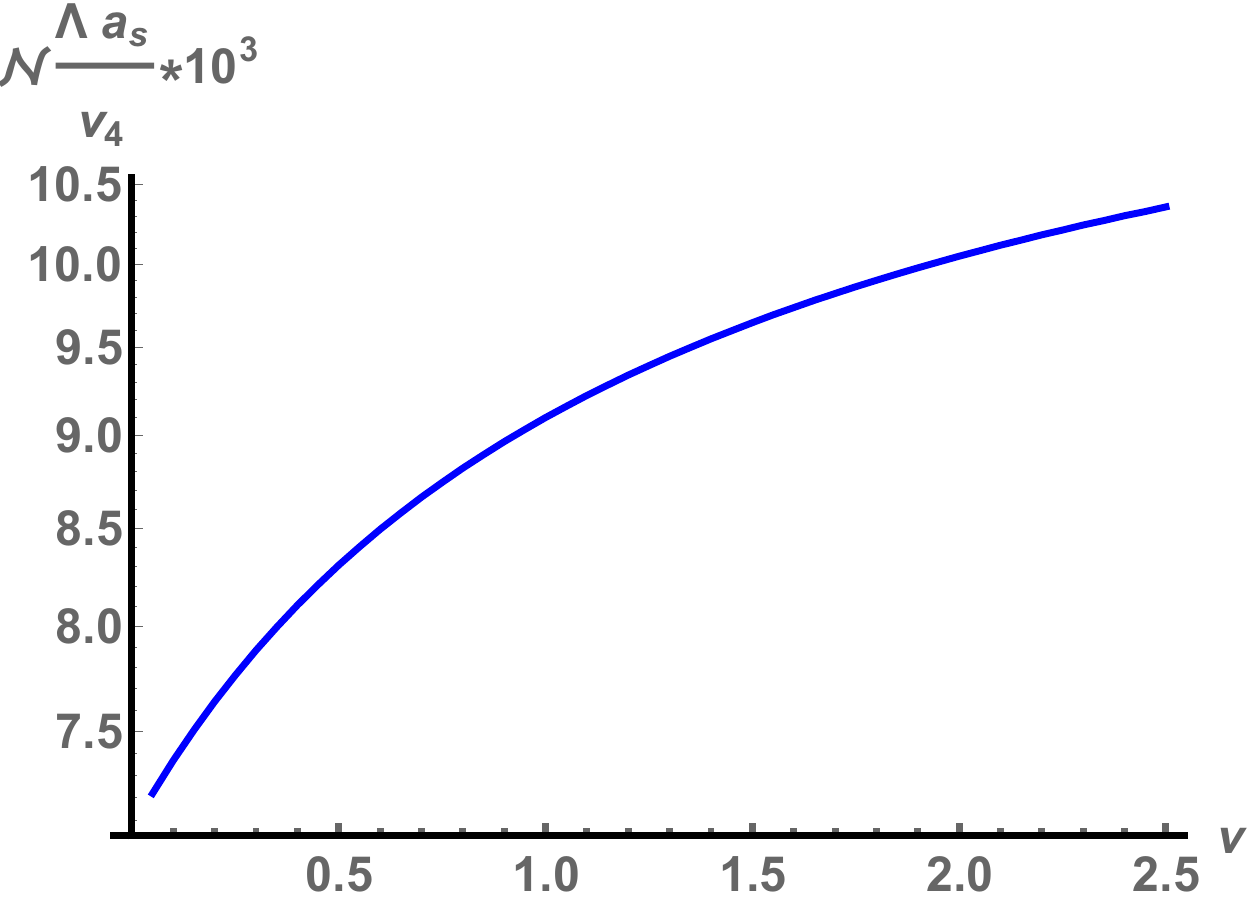}
\caption{Rescaled scattering length in units of the confinement scale. The horizontal axis is $\nu$, that up to a constant shift is the conformal dimension of the scalar operator that creates the particles involved in the scattering. In this case the scattering is for the particles of lowest mass at each value of the conformal dimension.}\label{fig:as0sol}
\end{subfigure}
\hspace{8pt}
\begin{subfigure}{0.45\textwidth}
\includegraphics[width=\textwidth]{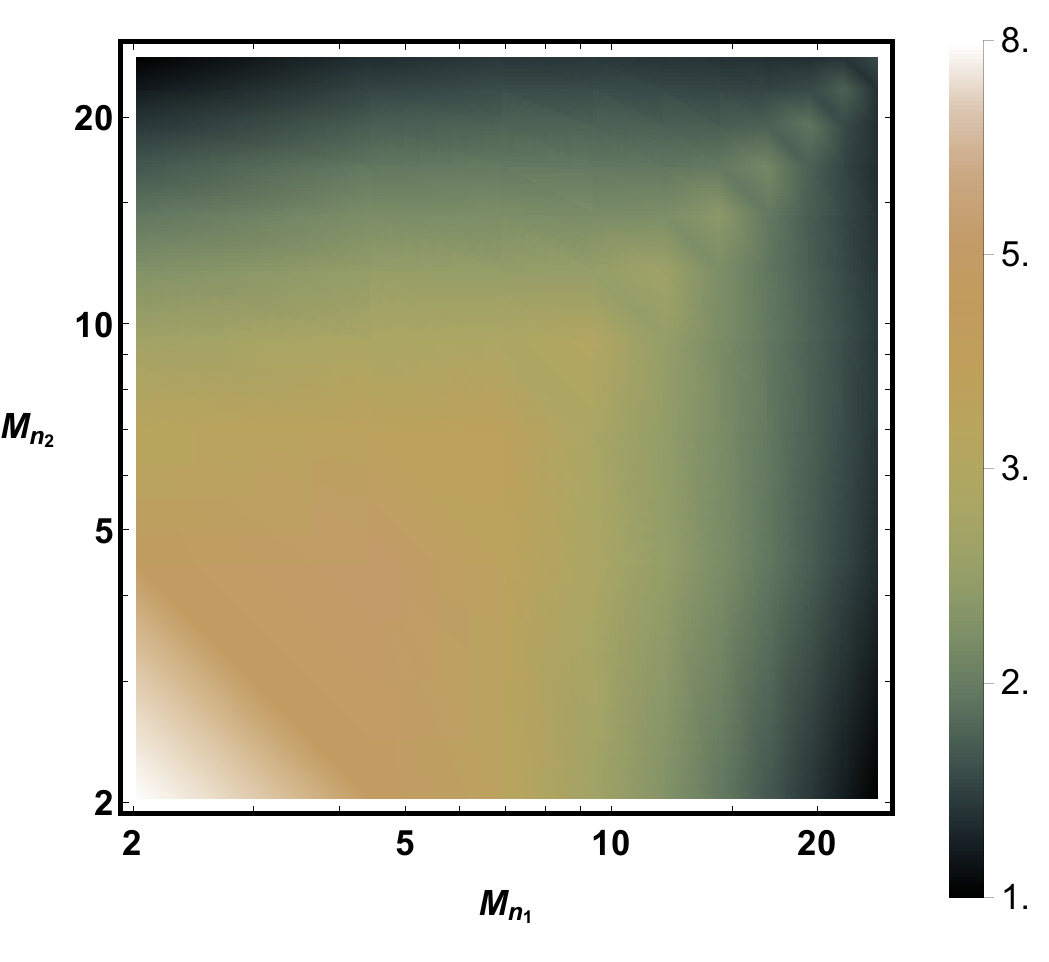}
\caption{Density plot of the scattering length $10^3\times {\cal{N}} a_s/v_4$ in units of the confinement scale. The vertical and horizontal axes scale logarithmically and indicate the masses of the particles involved in the scattering, in units of the confinement scale, for states created by a scalar operator of fixed conformal dimension (corresponding to $\nu=1/2$).}\label{fig:asmsol}
\end{subfigure}
\end{center}
\end{figure}

Using the values given in Table~\ref{table:masses}, the leading pole in the bulk-to-boundary propagator \eqref{eq:Ksol} is of the form \eqref{eq:poles}. In order to compute the residue numerically it is convenient to first evaluate numerically the limit\footnote{We explain how to generalize this calculation when $\nu$ is not a half-integer in Appendix~\ref{app:sol}.}
\be\label{eq:knnormal}
k_n=  \frac{1}{(2\nu)!}\lim_{M\to M_n}\; (M_n^2-M^2)\times\frac{\partial_z^{2\nu} \chi_{_M}(z)}{\chi_{_M}(0)}\Big|_{z=0} \ .
\ee
Then, the coefficient relating the residue to the normal mode is $c_n=k_n/\alpha_n$, and the residue of the two-point function of the dual operator is determined by
\be\label{eq:Z2sol}
Z_n=2\nu\, \cN\,k_n \ .
\ee
Some numerical results for $k_n$ are presented in Tables~\ref{table:masses} and \ref{table:excited} in Appendix~\ref{app:sol}.

The residue of the leading pole in the four-point function is given by \eqref{eq:res4}, evaluating the limit it becomes
\be\label{eq:Z4sol}
\cZ_{n_1; n_2,n_3,n_4}=-2\nu v_4\, \cN \,c_{n_2}c_{n_3}c_{n_4} \alpha_{n_1} k_{n_1}\kappa_{_{n_1},_{n_2},_{n_3},_{n_4}}\ ,
\ee
where the overlap $\kappa$ is defined as
\be
 \kappa_{_{n_1},_{n_2},_{n_3},_{n_4}} =\int_0^1 \frac{dz}{z^6} \prod_{i=1}^4 \varphi_{n_i}(z)\ .
\ee
We list some numerical values of the overlap in Appendix~\ref{app:sol}.

We proceed to compute the scattering length and the scattering amplitude for scatterers of equal and lowest mass. For this it is enough to use the formulas \eqref{eq:as} and \eqref{eq:amplit} plugging the values we have found for the residues \eqref{eq:Z2sol} and \eqref{eq:Z4sol}. The results for the lowest mass and different values of the conformal dimensions are represented in Figure~\ref{fig:as0sol}, while the results for scattering with different masses and fixed conformal dimensions can be found in Fig.~\ref{fig:asmsol}.

\section{Outlook}\label{sec:outlook}

In addition to its physical relevance, the similar behavior of the scattering length across different models suggests that this could be a good observable for further study via holographic duality. Available lattice data \cite{Briceno:2017max,Morningstar:2019jjz} can be used to check or calibrate holographic QCD models. In addition, comparison with the experimental data is possible and would be an interesting extension of the present work. 

A less direct application, but one where holography duality really stands out, is to use the information from the confining phase to make predictions about the properties of the finite temperature and non-zero density deconfined phases. That this is possible at all stems from the fact that the same classical action in the holographic dual describes all the phases. Contact terms in the action that enter in scattering amplitudes will show up in other quantities such as the equation of state, critical temperatures of phase transitions, and transport coefficients. For example, the coefficient of the quartic term in the bulk potential, $v_4$, is directly connected with the stiffness of the underlying equation of state for dense systems \cite{Hoyos:2016cob,Ecker:2017fyh}. 

Our analysis has been limited to a quartic contact interaction among scalars; see \cite{Hoyos:2019kzt} for a discussion on generalizations to higher order polynomial Lagrangian. It would be interesting to extend our framework to include the coupling to the metric, as well as to introduce other fields of different spin, in particular gauge fields. In general, there will be a non-trivial momentum dependence that would also be interesting to investigate.

\paragraph{Acknowledgements}

We would like to thank Massimo Bianchi and Maurizio Piai for useful comments. C.~H. is partially supported by the Spanish grant PGC2018-096894-B-100. C.~H. and D.~L. are partially supported by the Principado de Asturias through the grant FC-GRUPIN-IDI/2018/000174. N.~J. is supported in part by the Academy of Finland grant no. 1322307.

\appendix

\section{Normal modes and $\Delta<2$ in the $\cN=1^*$ SYM dual}\label{app:gppz}

In this appendix we will fill in some gaps in the calculations presented in Sec.~\ref{sec:GPPZ} in the main text.

\subsection{Normal modes}

The set of normalizable modes \eqref{eq:normalphi} maps to a set of Jacobi polynomials through the relation
\be
 P_{n}^{\left(\alpha,\beta\right)}\left(x\right)=\frac{\left(\alpha+1\right)_{n}}{n!}\,_{2}F_{1}\left(-n,1+\alpha+\beta+n;\alpha+1;\frac{1}{2}(1-x)\right) \ ,
\ee
where $(\alpha+1)_n=\Gamma(\alpha+n+1)/\Gamma(\alpha+1)$ is the Pochhammer's symbol. Then,
\be
 \phi_{M_n}(u)=\frac{1}{n+1} (1-u)^{\frac{2+\nu}{2}}P_{n}^{\left(1,\nu\right)}\left(1-2u\right) \ .
\ee
Jacobi polynomials satisfy the orthogonality condition
\be
 \int_{-1}^1 dx\,(1-x) (1+x)^\nu P_{n}^{\left(1,\nu\right)}(x) P_{m}^{\left(1,\nu\right)}(x)=\frac{2^{2+\nu}(n+1)}{(\nu+2n+2)(\nu+n+1)}\delta_{nm} \ .
\ee
Changing variables to $x=1-2u$, $0\leq u \leq 1$, the integral becomes ($\rho(u)=u/(2(1-u)^2)$)
\bea
& & \int_{-1}^1 dx\,(1-x) (1+x)^\nu P_{n}^{\left(1,\nu\right)}(x) P_{m}^{\left(1,\nu\right)}(x) \nonumber\\
&= & 2^{2+\nu}\int_0^1 du\, u (1-u)^\nu P_{n}^{\left(1,\nu\right)}(1-2u) P_{m}^{\left(1,\nu\right)}(1-2u)=\frac{2^{3+\nu}(n+1)^2}{\alpha_n^2}\int_0^1du\,\rho(u) \varphi_n(u)\varphi_m(u) \ .
\eea
Therefore, in order to satisfy the orthonormality condition, the coefficients $\alpha_n$ have to be fixed to
\be
\alpha_n^2=2(n+1)(\nu+n+1)(\nu+2n+2) \ .
\ee

\subsection{Alternative quantization}

In order to compute the two- and four-point functions using the approach in \cite{Hoyos:2019kzt} we first compute the one-point function as a function of the sources using the regularized canonical momentum associated to the radial evolution of the scalar
\be
 \pi_\phi^R=\sqrt{-g} g^{zz}\partial_z\phi+\frac{\delta S_{c.t.}}{\delta \phi} \ ,
\ee
where $S_{c.t.}$ is the counterterm action defined on a radial slice. For this discussion we will set the $AdS$ radius to unity $L=1$. The asymptotic expansion of the scalar field dual to an operator of dimension $\Delta_+=\frac{d}{2}+\nu$ takes the form
\be
 \phi \underset{z\to 0}{\longrightarrow} z^{\frac{d}{2}-\nu}+C_\nu z^{\frac{d}{2}+\nu} \ .
\ee
In this case, the one-point function of the dual operator is
\be
 \cN^{-1}\vev{\cO}_{\Delta_+}= \lim_{z\to 0} z^{\frac{d}{2}-\nu} \pi_\phi^R=2\nu C_\nu \ .
\ee
For $0<\nu<1$ there is an alternative quantization where the dual operator has dimension $\Delta_-=\frac{d}{2}-\nu$. In this case the asymptotic expansion of the scalar is
\be\label{eq:altexp}
 \phi \underset{z\to 0}{\longrightarrow} z^{\frac{d}{2}+\nu}+D_\nu z^{\frac{d}{2}-\nu} \ .
\ee
The leading contribution to the regularized momentum close to the boundary is
\be
 \pi_\phi^R \underset{z\to 0}{\longrightarrow} 2\nu z^{\nu-\frac{d}{2}} \ ,
\ee
and from this expression it follows that the expectation value of the dual operator is \cite{Papadimitriou:2007sj}
\be
 \cN^{-1}\vev{\cO}_{\Delta_-}= \lim_{z\to 0} \left(-2\nu z^{\nu-\frac{d}{2}}\phi \right) =-2\nu D_\nu \ .
\ee

Since the equation of motion for the scalar \eqref{eq:eomGPPZ} does not depend on the sign of $\nu$, symmetry under a sign flip $\nu\to -\nu$ is expected. Indeed, using Euler's relation
\be
 \,_{2}F_{1}\left(a,b;c;u\right)=(1-u)^{c-a-b}\,_{2}F_{1}\left(c-a,c-b;c;u\right)=(1-u)^{c-a-b}\,_{2}F_{1}\left(c-b,c-a;c;u\right) \ ,
\ee
the solution for the scalar can also be written as
\be
 \phi_M(u)=(1-u)^{\frac{2-\nu}{2}}\,_{2}F_{1}\left(1-\frac{\nu}{2}+\frac{1}{2}\sqrt{\nu^{2}+M^{2}-4},1-\frac{\nu}{2}-\frac{1}{2}\sqrt{\nu^{2}+M^{2}-4};2;u\right) \ .
\ee
The bulk-to-boundary propagator is then the same as in \eqref{eq:KGPPZ} with the replacement $\nu\to -\nu$. This implies that the coefficient in \eqref{eq:altexp} is $D_\nu=C_{-\nu}$ and then
\be
 \vev{\cO}_{\Delta_-}=\vev{\cO}_{\Delta_+}\Big|_{\nu\to -\nu} \ .
\ee
Since the two- and four-point functions and their residues are all derived from the one-point function, it follows that the scattering length and amplitude of states created by operators of dimension $\Delta_-=2-\nu$, $0<\nu<1$ can be obtained simply by continuing the results for $\Delta_+=2+\nu$ to negative values $-1<\nu<0$.

\section{Numerical solutions in the $AdS_6$ soliton}\label{app:sol}

In this appendix we will detail some steps for interested reader to reproduce the numerical data. In order to find the solution we use numerical shooting, starting with a regular solution at the origin $z=1$ that is expanded in a power series $\chi_t(z)$ with a fixed value $\chi_t(1)=1$ at the tip of the $AdS$ soliton
\be 
 \chi_t(z)=1+\sum_{n=1}^{N_t} a_n(1-z)^n\ . 
\ee
Here we will consider $N_t=8$, which is large enough to have reliable results. The value of the coefficients for the first terms in the series are
\bea
a_1 & = & \frac{1}{20} \left((5-2 \nu )^2-4 M^2\right)\\
a_2 & = & \frac{1}{1600}\left((5-2 \nu )^2 (4 (\nu -15) \nu +65)+16 M^4-8 (4 (\nu -10) \nu +75) M^2\right) \\
a_3 & = & \frac{1}{288000}\left((5-2 \nu )^2 (8 \nu  (\nu  (2 (\nu -40) \nu +815)-2100)+7425)-64 M^6\right. \\
 & & \left.+16 (12 (\nu -15) \nu +455) M^4-4 (24 \nu  (\nu (2 (\nu -30) \nu +485)-1425)+30175) M^2\right) \ .
\eea
This series is used to give boundary conditions to the numerical solution $\chi_N(z)$ at $z=1-\epsilon$, where we take $\epsilon=10^{-6}$. So we fix
\be
 \chi_N(1-\epsilon)=\chi_t(1-\epsilon)\ , \ \chi_N'(1-\epsilon)=\chi_t'(1-\epsilon) \ .
\ee
The numerical solution and its derivative up to order $2\nu$ can be found, {\emph{e.g.}}, using NDSolve in Mathematica with these boundary conditions, in the interval $z\in [\epsilon,1-\epsilon]$. The boundary value is taken to be approximately the value of the numerical solution at $z=\epsilon$:
\be
 \chi_{_M}(0)\approx \chi_N(\epsilon) \ .
\ee
The numerical value of the derivative at the boundary is also approximated by the value of the numerical solution at $z=\epsilon$:
\be
 \partial_z^{2\nu}\chi(0) \approx \partial_z^{2\nu}\chi_N(\epsilon) \ .
\ee

If $\nu$ is not half-integer, then we need to be more cautious. In this case, we first do a change of variables in the radial coordinate $z=u^{1/(2\nu)}$ and proceed with the same shooting method. The equation \eqref{eq:knnormal} is replaced by
\be
 k_n= \lim_{M\to M_n}\; (M_n^2-M^2)\times\frac{\partial_u \chi_{_M}(u)}{\chi_{_M}(0)}\Big|_{u=0} \ .
\ee
Numerically, this is evaluated using
\be
 \partial_u\chi(0) \approx \partial_u\chi_N(u=\epsilon^{2\nu}) \ .
\ee
What remains to be done in order to find the spectrum of normal modes is to solve for $\chi_{_M}(0)=0$ for fixed $M$. The value $M$ is tuned until a zero is found, an analysis that can be effectively performed using Newton's method.
To fix the coefficients $\alpha_n$ we evaluate numerically the following integrals 
\be
\alpha_n^{-2} \approx \int_{\epsilon}^{1-\epsilon} dz\, z^{1-2\nu} \chi_N(z)^2\Big|_{M=M_n}\ .
\ee
Finally, we compute the overlaps in a similar way, by performing the following integral numerically
\be
\frac{\kappa_{n_1,n_2,n_3,n_4}}{\alpha_{n_1}\alpha_{n_2}\alpha_{n_3}\alpha_{n_4}}\approx \int_{\epsilon}^{1-\epsilon} dz\, z^{4-4\nu} \prod_{i=1}^4 \chi_N(z)\Big|_{M=M_{n_i}}\ .
\ee
We have collected some numerical values in Table~\ref{table:masses} for the lowest mode and in Table~\ref{table:excited} for excited states for fixed $\nu$ that we obtain from numerical calculation.

\begin{table}
 \begin{center}
	\begin{tabularx}{0.58\textwidth}{cYYYYY}
	\hline
		$\nu$ & 1/2 &  1 & 3/2 & 2 & 5/2 \\
	\hline
	\hline	$M_0$  & 2.02 & 2.54 & 3.05 & 3.56 & 4.06  \\  
	\hline	$M_{\rm WKB}$  & 1.77 & 2.51 & 3.07 & 3.54 & 3.96 \\ 
	\hline $\alpha_0$ & 1.36 & 1.57 & 1.76 & 1.93  & 2.09 \\ 
    \hline $k_0$ & 6.06 & 20.2 & 47.2 & 94.2 & 171. \\ 
    \hline $\kappa_{0,0,0,0}$ & 0.85 & 1.16 & 1.48 & 1.80 & 2.11 \\  
	\hline
	\end{tabularx}
	\caption{Numerical values obtained from shooting for lowest modes $n=0$. Notice that the WKB approximation detailed in Appendix~\ref{app:wkb} gives very accurate results.}\label{table:masses}
\end{center}
	\end{table}

\begin{table}
 \begin{center}
	\begin{tabularx}{0.88\textwidth}{cYYYYYYYYYY}
	\hline
		$n$ & 0 & 1 & 2 & 3 & 4 & 5 & 6 & 7 & 8 & 9 \\
	\hline
	\hline	$M_n$  & 2.02  & 4.45  & 6.93 & 9.43 & 11.93 & 14.43 & 16.93 & 19.43 & 21.94 & 24.44 \\
	\hline	$M_{\rm WKB}$  & 1.77  & 4.34  & 6.86 & 9.38 & 11.89 & 14.39 & 16.90 & 19.41 & 21.92 & 24.42 \\  
    \hline  $\alpha_n$ & 1.36  & 2.10 & 2.63 & 3.07 & 3.45 & 3.80 & 4.12 & 4.41 & 4.69 & 4.95 \\ 
    \hline $k_n$ & 6.06  & 31.1  & 76.2 & 141. & 226. & 332. & 457. & 602. & 767. & 953. \\ 
    \hline
	\end{tabularx}
	\caption{Numerical values obtained from shooting for excited modes for fixed $\nu=1/2$. Notice that the WKB approximation detailed in Appendix~\ref{app:wkb} gives very accurate results and becomes increasingly better for higher modes as expected.}\label{table:excited}
\end{center}
	\end{table}

\section{WKB approximation}\label{app:wkb}

Following the method developed in \cite{Russo:1998by} we will compute the masses of normal modes using the WKB approximation. First, we will write the equations of motion in the following form
\be\label{eq:eqwkb}
 \partial_x( f(x) y(x))+\left( M^2 h(x)+p(x)\right) y(x)=0 \ .
\ee
The value of the masses can be obtained from the behavior of these functions close to the origin $x\to 1$ and the boundary $x\to \infty$. Close to the origin $x\to 1$,
\be\label{eq:exporigin}
 f\approx f_1(x-1)^{s_1} \ , \ h\approx h_1(x-1)^{s_2}\ , \ p\approx p_1(x-1)^{s_3} \ ,
\ee
whereas close to the boundary $x\to \infty$,
\be\label{eq:expboundary}
 f\approx f_2 x^{r_1}\ ,  \ h\approx h_2 x^{r_2}\ , \ p\approx p_2 x^{r_3} \ .
\ee
The WKB approximation of the masses is given by the formula
\be\label{eq:WKBmass}
 M_n^2=\frac{\pi^2}{\xi^2}(n+1)\left(n+\frac{\alpha_2}{\alpha_1}+\frac{\beta_2}{\beta_1}\right)+O(n^0)\ , \ n\geq 0 \ .
\ee
The quantities appearing in this expression are, an integral setting the scale
\be\label{eq:xi}
 \xi=\int_1^\infty dx \sqrt{\frac{h}{f}} \ ,
\ee
and the following combination of exponents and coefficients of the asymptotic expansions
\be\label{eq:wkbcoef1}
 \alpha_1=s_2-s_1+2, \ \beta_1=r_1-r_2-2 \ , 
\ee
and
\be\label{eq:wkbcoef2}
\alpha_2=\left\{ \begin{array}{ll} |s_1-1|, & s_3-s_1+2\neq 0 \\ \sqrt{(s_1-1)^2-4\frac{p_1}{f_1}}, & s_3-s_1+2=0\end{array} \right. \ , \ \beta_2=\left\{ \begin{array}{ll}  |r_1-1| & r_1-r_3-2\neq 0 \\ \sqrt{(r_1-1)^2-4\frac{p_2}{f_2}}, & r_1-r_3-2=0 \end{array} \right.\ .
\ee

\subsection{$\cN=1^*$ SYM}
Let us now specify to the first case studied in this paper in Sec.~\ref{sec:GPPZ}. While we were able to obtain the mass spectrum analytically, it is interesting to compare how close the WKB approximation is with the exact values.
By comparing with the exact result, we will have an estimate of the error in the approximation for cases where the exact result is unknown. Starting with \eqref{eq:eomGPPZ}, we make a change variables $u=1-1/x$ as well as a field redefinition 
\be
 \phi(x)=x^{-1/2}(x-1)^{-1} y(x) \ .
\ee
Then, the equation for $y(x)$ has the form \eqref{eq:eqwkb} with
\be
 f(x)=1 \ , \ h(x)=\frac{1}{4 x^2(x-1)} \ , \ p(x)=\frac{x-5+\nu^2(1-x)}{4 x^2(x-1)} \ .
\ee
The asymptotic form of these functions close to the origin $x\to 1$ is as in \eqref{eq:exporigin} with
\be
f_1=1\ , \ s_1=0\ ;  \ h_1=\frac{1}{4}\ , \ s_2=-1\ ; \ p_1=-1\ , \ s_3=-1\ .
\ee
Close to the boundary $x\to \infty$ the expansion is as in \eqref{eq:expboundary} with
\be
f_2=1\ , \ r_1=0\ ; \ h_2=\frac{1}{4}\ ,\  r_2=-3\ ; \ p_2=\frac{1-\nu^2}{4}\ , \ r_3=-2\ .
\ee
We will plug these expressions in \eqref{eq:xi}, \eqref{eq:wkbcoef1}, and \eqref{eq:wkbcoef2}, taking into account that $r_1-r_3-2=0$. This yields
\be
 \xi=\frac{\pi}{2} \ , \ \alpha_1=1\ , \ \beta_1=1 \ , \ \alpha_2=1 \ , \ \beta_2=\nu \ .
\ee
Using these values in the mass formula \eqref{eq:WKBmass}, the WKB approximation gives
\be
 M_n^2=4(n^2+(2+\nu)n+(1+\nu))+O(n^0)\ , \ n\geq 0 \ .
\ee
Comparing with the exact formula \eqref{eq:MnGPPZ} we see that the $O(n^0)$ correction is a $\nu$-independent term
\be
 (M_n^2)_{\rm exact}-(M_n^2)_{\rm WKB}=4\ . 
\ee

\subsection{Non-supersymmetric theory}

Let us now proceed with comparing the mass spectra for the non-supersymmetric case studied numerically in Sec.\ref{sec:AdS6}.
In this case we do not have an analytic expression for the masses, but we can compare the results from the WKB approximation with those from the numerical computation. Starting with the equation of motion \eqref{eq:EOMsol}, we make a change of variables $z=1/x$ and simple rename the field $\phi(x)=y(x)$. This results in an equation of the form \eqref{eq:eqwkb} with
\be
 f(x)=x^6-x\ ,\ h(x)=x^2\ , \ p(x)=\left( \frac{25}{4}-\nu^2\right)x^4 \ .
\ee
The asymptotic form of these functions close to the origin $x\to 1$ is as in \eqref{eq:exporigin} with
\be
f_1=5\ , \ s_1=1\ ;  \ h_1=1\ ,\ s_2=0\ ; \ p_1=\frac{25}{4}-\nu^2\ ,\  s_3=0\ .
\ee
Close to the boundary $x\to \infty$ the expansion is as in \eqref{eq:expboundary} with
\be
f_2=1\ , \ r_1=6\ ;  \ h_2=1\ ,\ r_2=2\ ; \ p_2=\frac{25}{4}-\nu^2\ ,\ r_3=4 \ .
\ee
We will plug these expressions in \eqref{eq:xi}, \eqref{eq:wkbcoef1}, and \eqref{eq:wkbcoef2}, taking into account that $r_1-r_3-2=0$. This yields
\be
 \xi=\frac{\sqrt{\pi}\Gamma\left(6/5\right)}{\Gamma\left(7/10\right)} \ , \ \alpha_1=1\ , \ \beta_1=2\ ,  \ \alpha_2=0\ , \ \beta_2=2\nu \ .
\ee
Using these values in the mass formula \eqref{eq:WKBmass}, the WKB approximation gives
\be
 M_n^2=\frac{\pi^2}{\xi^2}(n+1)(n+\nu)+O(n^0) \ , \ n\geq 0 \ .
\ee
We note that the WKB approximation compares really well with the numerical values, even for lowest lying modes.

\bibliographystyle{JHEP}
\bibliography{refsscat}

\end{document}